\documentstyle[psfig,12pt]{article}

\def\beq{\begin{equation}}
\def\eeq{\end{equation}}

\def\be{\begin{equation}}
\def\bea{\begin{eqnarray}}
\def\ee{\end{equation}}
\def\eea{\end{eqnarray}}

\begin{document}


\begin{titlepage}

\begin{flushright} UMN-D-02-2 \end{flushright}

\begin{centering}

\vspace*{3cm}

{\Large\bf Anomalously light states in super-Yang--Mills Chern--Simons
theory}

\vspace*{1.5cm}

{\bf J.R.~Hiller,$^a$ S.S.~Pinsky,$^b$ U.~Trittmann$^b$}
\vspace*{0.5cm}

$^a$
{\sl Department of Physics \\
University of Minnesota Duluth\\
Duluth, MN  55812}

\vspace*{0.5cm}

$^b$
{\sl Department of Physics\\
Ohio State University\\
Columbus, OH 43210, USA}

\vspace*{1cm}

\vspace*{1cm}

\begin{abstract}
Inspired by our previous finding that supersymmetric 
Yang--Mills-Chern--Simons (SYM-CS) theory dimensionally reduced to 
1+1 dimensions possesses approximate Bogomol'nyi--Prasad--Sommerfield (BPS)
states, we study the analogous phenomenon in the three-dimensional theory. 
Approximate BPS states in two dimensions have masses which are nearly 
independent of the Yang-Mills coupling and proportional to their average 
number of partons. These states are a reflection of the exactly massless 
BPS states of the underlying pure SYM theory.  In three dimensions
we find that this mechanism leads to anomalously light bound states. While 
the mass scale is still proportional to the average number of partons times 
the square of the CS coupling, the average number of partons in these bound 
states changes with the Yang--Mills coupling. Therefore, the masses of these
states are not independent of the coupling. Our numerical calculations are done using supersymmetric discrete light-cone quantization (SDLCQ).   
\end{abstract}
\end{centering}

\vfill

\end{titlepage}
\newpage

\section{Introduction}

It is extremely interesting to look for mechanisms in quantum field theory 
that give rise to anomalously light bound states, since these are the states 
we first see as we investigate new phenomena. Of course, spontaneous chiral 
symmetry breaking is the most profound example of these mechanisms. The 
underlying chiral symmetry of QCD gives rise to exactly massless bound states, 
the Goldstone bosons, which acquire masses from small mass terms for the
constituent quarks which in turn break the symmetry. 

What we are suggesting here is that there may be a similar mechanism at work in
supersymmetric theories. ${\cal N}=1$ pure supersymmetric Yang--Mills
(SYM) theory in 1+1 and 2+1 dimensions has massless BPS 
states~\cite{hpt2001,hpt2001b}. Recently~\cite{CSSYM1+1,BPS1+1} we created 
an effective mass for the constituents of a SYM theory in 1+1 dimensions by
adding a dimensionally reduced Chern--Simons (CS) term~\cite{dunne} to the
theory. Although such a term does not break the
supersymmetry, the theory with a CS term does not have BPS states. We found,
however, that this theory has approximate BPS states, i.e.~states whose masses 
are nearly independent of the Yang--Mills (YM) coupling. Furthermore, at large
YM coupling the masses of these states are equal to the square of the CS 
coupling times the average number of particles. 

In this paper we will 
investigate the analog of these states in the (2+1)-dimensional theory. We will
show that in 2+1 dimensions these states persist, although with some significant
changes. We will see that in 2+1 dimensions the masses of these states are not
independent of the YM coupling as they are in 1+1 dimensions. This is 
primarily because the average number of particles in these states grows with 
the coupling. However, the masses of these states do not grow as fast as the
masses of the other bound states, and, therefore, at strong coupling these 
states are anomalously light.

The numerical method we will use is SDLCQ (Supersymmetric Discrete Light-Cone
Quantization).  This is a numerical method that can be used to solve any theory
with enough supersymmetry to be finite.  The central point of this method is 
that using DLCQ~\cite{pb85,bpp98} we can construct a finite dimensional
representation of the superalgebra~\cite{sakai95}.  From this representation 
of the superalgebra,  we construct a finite-dimensional Hamiltonian which we
diagonalize  numerically. We repeat the process for larger and
larger representations and extrapolate the solutions to the continuum.  

In constructing the discrete approximation we drop the
longitudinal zero-momentum mode.  For some discussion of
dynamical and constrained zero modes, see the review~\cite{bpp98} 
and previous work~\cite{alpt98}.
Inclusion of these modes would be ideal, but the techniques
required to include them in a numerical calculation
have proved to be difficult to develop, particularly because
of nonlinearities.   For DLCQ calculations that can be 
compared with exact solutions, the exclusion of
zero modes does not affect the massive spectrum~\cite{bpp98}.
In scalar theories it has been known for some time that 
constrained zero modes can give rise to dynamical symmetry 
breaking~\cite{bpp98}, and work continues on the role of zero modes 
and near zero modes in these theories~\cite{thorn}.
Dropping the zero modes in CS theory does result in the loss of many 
of the interesting aspects of CS theory, most notably the quantization 
of the CS coupling. 

One particularly interesting property that {\em will} be preserved is 
the fact that the CS term simulates a mass for the theory. This property
is interesting because reduced
${\cal N}=1$ SYM theories are very stringy theories.  The low-mass states are
dominated by Fock states with many constituents, and as the size of the
superalgebraic representation is increased, states with lower masses and
more constituents appear~\cite{hpt2001,hpt2001b,alpt98,Lunin:2001im,%
Pinsky:2000rn,Hiller:2000nf,Haney:2000tk,Lunin:1999ib,Antonuccio:1999zu,%
Antonuccio:1998mq,Antonuccio:1998tm,Antonuccio:1998jg,Antonuccio:1998kz}. 
The connection between string theory and supersymmetric gauge theory leads 
one to expect this type of behavior. However, these gauge theories are not
very QCD-like, and ultimately one might like to make a connection with the
low-mass spectrum observed in nature.  The effective mass supplied by
the CS term makes the theory more like QCD.

The present Hamiltonian formalism provides us with the wave functions 
from which we construct the structure functions of the mass eigenstates. 
We will see that the structure functions of the approximate BPS
states have unique shapes which might help to single them out experimentally. 

The paper is structured as follows.  In Sec.~\ref{sec:SuperCS}
we provide a summary of SYM-CS theory; much of this is given in~\cite{CSSYM1+1}
but is repeated here for completeness.  Our numerical techniques are described
in Sec.~\ref{sec:numerical}.  An analysis of what should be expected for SYM-CS
eigenstates is given in Sec.~\ref{sec:stage}, and our results are presented
and discussed in Sec.~\ref{sec:results}.  Section~\ref{sec:Outlook} contains a
summary of our conclusions and some discussion of interesting further investigations.

\section{Supersymmetric Chern--Simons theory} 
\label{sec:SuperCS}

The Lagrangian of (2+1)-dimensional ${\cal N}=1$ SYM-CS theory is 
\begin{equation}
{\cal L}={\rm Tr}(-\frac{1}{4}{\cal L}_{\rm YM}+i{\cal L}_{\rm F}
+\frac{\kappa}{2}{\cal L}_{\rm CS}),\label{Lagrangian}
\end{equation}
where $\kappa$ is the CS coupling and
\begin{eqnarray}
{\cal L}_{\rm YM}&=&F_{\mu\nu}F^{\mu\nu}\,, \\
{\cal L}_{\rm F}&=&\bar{\Psi}\gamma_{\mu}D^{\mu}\Psi\,, \\
{\cal L}_{\rm CS}&=&\epsilon^{\mu\nu\lambda}\left(A_{\mu}
\partial_{\nu}A_{\lambda}+\frac{2i}{3}gA_\mu A_\nu A_\lambda \right)
+2\bar{\Psi}\Psi\,. \label{eq:CSLagrangian}
\end{eqnarray}
The two components of the spinor $\Psi=2^{-1/4}({\psi \atop \chi})$ 
are in the adjoint representation of $U(N_c)$ or $SU(N_c)$. 
We will work in the large-$N_c$ limit.  The field strength and the 
covariant derivative are
\begin{equation}
F_{\mu\nu}=\partial_{\mu}A_{\nu}-\partial_{\nu}A_{\mu}
              +ig[A_{\mu},A_{\nu}]\,, \quad \quad 
D_{\mu}=\partial_{\mu}+ig[A_{\mu},\quad]\,.
\end{equation} 
The supersymmetric variations of the fields leads to the supercurrent 
$Q^{(\mu)}$ in the usual manner via  
\begin{equation}
\delta{\cal L}=\bar{\epsilon}\partial_{\mu}Q^{(\mu)}.
\label{LQ}
\end{equation}
Light-cone coordinates in 2+1 dimensions are $(x^+,x^-,x^\perp)$ where 
$x^+=x_-$ is the light-cone time and $x^\perp=-x_\perp$. The totally 
anti-symmetric tensor is defined by $\epsilon^{+-2}=-1$.
The variations of the three parts of the Lagrangian in Eq.~(\ref{Lagrangian}) determine the (`chiral') components $Q^\pm$ of the supercharge
via Eq.~(\ref{LQ}) to be
\begin{equation}
\int d^2x Q^{(+)}=\left( {Q^+ \atop Q^-}\right)
   = -\frac{1}{2}\int d^2x\,
\epsilon^{\alpha\beta\lambda}\gamma_\lambda\gamma^{+}\Psi F_{\alpha\beta}\,.
\end{equation}
Explicitly they are 
\begin{eqnarray} \label{supercharges}
Q^-&=&-i2^{3/4}\int d^2x\,
\psi\left(\partial^+ A^- -\partial^-A^+ + ig[A^+,A^-]\right)\,,
\nonumber \\
Q^+&=&-i2^{5/4}\int d^2x\,
\psi\left(\partial^+ A^\perp -\partial^2 A^+ + ig[A^+,A^\perp]\right)\,.
\end{eqnarray}
One can convince oneself by calculating the energy-momentum tensor
$T^{\mu\nu}$ that the supercharge fulfills the supersymmetry algebra
\begin{equation}
\{Q^\pm,Q^\pm\}=2\sqrt2 P^\pm\,, \qquad 
\{Q^+,Q^-\}=-4P^\perp\,.
\end{equation}

In order to express the supercharge in terms of the physical degrees of
freedom, we have to use equations of motion, some of which are
constraint equations. The equations of motion for the gauge fields are
\begin{equation}
-D_\nu F^{\nu\alpha}=\frac{\kappa}{2}\epsilon^{\alpha\nu\lambda}
F_{\nu\lambda}+2g\bar{\Psi}\gamma^\alpha\Psi\,.
\end{equation}
For $\alpha=+$ this is a constraint for $A^-$.
In light-cone gauge, $A^+=0$, the constraint reduces to
\begin{equation} \label{eq:A-Constraint}
D_-A^-=\frac{1}{D_-}[(\kappa-D_2)D_-A^\perp+2g\bar{\Psi}\gamma^+\Psi]\,.
\end{equation}
The equation of motion for the fermion is
$\gamma^\mu D_\mu \Psi=-i\kappa\Psi$.
Expressing everything in terms of $\psi$ and $\chi$ leads to the 
equations of motion
\begin{eqnarray}
\sqrt{2}D_+\psi&=&(D_2+\kappa)\chi\,,  \\
\sqrt{2}D_-\chi&=&(D_2-\kappa)\psi\,, \label{eq:chiConstraint}
\end{eqnarray}
the second of which is a constraint equation.  The constraint equations,
(\ref{eq:A-Constraint}) and (\ref{eq:chiConstraint}),
are used to eliminate $A^-$ and $\chi$.

We compactify in the $x^\perp$ direction with compactification length
$L$, meaning that transverse momentum modes are summed over a discrete
set of values $2\pi n^\perp/L$.  In order to have a finite 
matrix representation for the eigenvalue problem, we must 
truncate these sums at some fixed integers $\pm T$.  The
value of $T$ defines a physical transverse cutoff
$\Lambda_\perp=2\pi T/L$; however, given this definition,
$T$ can be viewed as a measure of transverse resolution
at fixed $\Lambda_\perp$.  

At large $N_c$ the expansions of the field operators $\phi\equiv A_\perp$ 
and $\psi$, in terms of creation and annihilation operators for the Fock 
basis, are
\bea
\lefteqn{
\phi_{ij}(0,x^-,x_\perp) =} & & \nonumber \\
& &
\frac{1}{\sqrt{2\pi L}}\sum_{n^{\perp} = -\infty}^{\infty}
\int_0^\infty
         \frac{dk^+}{\sqrt{2k^+}}\left[
         a_{ij}(k^+,n^{\perp})e^{-{\rm i}k^+x^- +{\rm i}
\frac{2 \pi n^{\perp}}{L} x_\perp}+
         a^\dagger_{ji}(k^+,n^{\perp})e^{{\rm i}k^+x^- -
{\rm i}\frac{2 \pi n^{\perp}}{L}  x_\perp}\right]\,,
\nonumber\\
\lefteqn{
\psi_{ij}(0,x^-,x_\perp) =} & & \nonumber \\
& & \frac{1}{2\sqrt{\pi L}}\sum_{n^{\perp}=-\infty}^{\infty}\int_0^\infty
         dk^+\left[b_{ij}(k^+,n^{\perp})e^{-{\rm i}k^+x^- +
{\rm i}\frac{2 \pi n^{\perp}}{L} x_\perp}+
         b^\dagger_{ji}(k^+,n^\perp)e^{{\rm i}k^+x^- -{\rm i}
\frac{2 \pi n^{\perp}}{L} x_\perp}\right]\,.
\nonumber
\eea
From the field \mbox{(anti-)}commutators one finds
\begin{equation}
\left[a_{ij}(p^+,n_\perp),a^\dagger_{lk}(q^+,m_\perp)\right]=
\left\{b_{ij}(p^+,n_\perp),b^\dagger_{lk}(q^+,m_\perp)\right\}=
\delta(p^+ -q^+)\delta_{n_\perp,m_\perp}\delta_{il}\delta_{jk}\,.
\end{equation}
The truncated supercharge $Q^-$ can be written as
\begin{equation}
\label{Qminus}
Q^- = g Q^-_{\rm SYM}(T) + Q_\perp (T) + i\kappa Q^-_{\rm CS}(T)\,,
\end{equation}
where
\begin{equation}
\label{Qperp}
Q^-_\perp(T)=\frac{2^{3/4}\pi {\rm i}}{L}\sum_{|n^\perp|\le T}\int_0^\infty dk
\frac{n^\perp}{\sqrt{k}}\left[
a_{ij}^\dagger(k,n^\perp) b_{ij}(k,n^\perp)-
b_{ij}^\dagger(k,n^\perp) a_{ij}(k,n^\perp)\right]\,,
\end{equation}
\begin{equation}
\label{QCS}
Q^-_{\rm CS}(T)=2^{-1/4}{\rm i}\sum_{|n^\perp|\le T}\int_0^\infty dk
\frac{1}{\sqrt{k}}\left[
a_{ij}^\dagger(k,n^\perp) b_{ij}(k,n^\perp)+b_{ij}^\dagger(k,n^\perp)
a_{ij}(k,n^\perp)\right]\,,
\end{equation}
and
\begin{eqnarray}
\label{QSYM}
&&Q^-_{\rm SYM}(T)=
{{\rm i} 2^{-5/4} \over \sqrt{L\pi}}\sum_{|n_i^\perp|\le T}\int_0^\infty dk_1dk_2dk_3
\delta(k_1+k_2-k_3) \delta_{n^\perp_1+n^\perp_2,n^\perp_3}
\left\{2\left({ 1\over k_1}+{1 \over k_2}-{1\over k_3}\right)\right.\nonumber\\
&&\qquad \qquad \times\left[b_{ik}^\dagger(k_1,n^\perp_1) b_{kj}^\dagger(k_2,n^\perp_2)
b_{ij}(k_3,n^\perp_3)
+b_{ij}^\dagger(k_3,n^\perp_3) b_{ik}(k_1,n^\perp_1) 
      b_{kj}(k_2,n^\perp_2)]\right]        \nonumber \\
    && +{k_2-k_1  \over  k_3\sqrt{k_1 k_2}} 
\left[a_{ik}^\dagger(k_1,n^\perp_1) a_{kj}^\dagger(k_2,n^\perp_2)
b_{ij}(k_3,n^\perp_3)
-b_{ij}^\dagger(k_3,n^\perp_3)a_{ik}(k_1,n^\perp_1)
a_{kj}(k_2,n^\perp_2) \right]\nonumber\\
&&+{k_1+k_3  \over k_2 \sqrt{k_1 k_3}}
\left[a_{ik}^\dagger(k_3,n^\perp_3) a_{kj}(k_1,n^\perp_1) b_{ij}(k_2,n^\perp_2)
-a_{ik}^\dagger(k_1,n^\perp_1) b_{kj}^\dagger(k_2,n^\perp_2)
a_{ij}(k_3,n^\perp_3) \right]\nonumber\\
&&+{k_2+k_3  \over k_1\sqrt{k_2 k_3}}
\left[b_{ik}^\dagger(k_1,n^\perp_1) a_{kj}^\dagger(k_2,n^\perp_2)
a_{ij}(k_3,n^\perp_3)
-a_{ij}^\dagger(k_3,n^\perp_3)b_{ik}(k_1,n^\perp_1) 
      a_{kj}(k_2,n^\perp_2) \right] \left. \frac{}{}\right\}\,. \nonumber\\
\end{eqnarray}
The symmetric truncation with respect to positive
and negative modes aids in retaining a reflection parity
symmetry in the states. The light-cone energy is $(k^2_\perp + m^2)/k^+$,
and $k_\perp$ behaves like a mass.  Here we see that $\kappa$ appears in a 
very similar way to $k_\perp$ and therefore behaves in many ways like a mass.

When comparing the $\perp$ and CS contributions to the supercharge, we
see that we have a relative $i$ between them. Thus the usual eigenvalue
problem
\begin{equation}
2P^+P^-|\varphi\rangle=\sqrt{2}P^+(Q^-)^2|\varphi\rangle=
\sqrt{2}P^+(gQ^-_{\rm SYM}+Q_\perp^-
      +i\kappa Q^-_{\rm CS})^2|\varphi\rangle=M^2_n|\varphi\rangle
\label{EVP}
\end{equation}
has to be solved by using fully complex methods. 

We retain\footnote{We note that the CS term breaks transverse parity.}
the $S$-symmetry, which is associated with the orientation of the 
large-$N_c$ string of partons in a state~\cite{kutasov93}. In a 
(1+1)-dimensional model this orientation parity is usually referred as a $Z_2$
symmetry, and we will follow that here.  It gives a sign when the color
indices are permuted
\begin{equation}\label{Z2}
Z_2 : a_{ij}(k,n^\perp)\rightarrow -a_{ji}(k,n^\perp)\,, \qquad
      b_{ij}(k,n^\perp)\rightarrow -b_{ji}(k,n^\perp)\,.
\end{equation}
We will use this symmetry to reduce the Hamiltonian matrix size and
hence the numerical effort. All of our states will be labeled by the
$Z_2$ sector in which they appear. We will not attempt to label the states 
by their normal parity; in light-cone coordinates this is only an approximate 
symmetry. Such a labeling could be done in an approximate way,
as was shown by Hornbostel~\cite{horn}, and might be useful for 
comparison purposes if at some point there are results from 
lattice simulations of the present theory. 
 
\section{Numerical Methods}
\label{sec:numerical}

We convert the mass eigenvalue problem $2P^+P^-|M\rangle = M^2 |M\rangle$ 
to a matrix eigenvalue problem by introducing a discrete $P^-$ in
a basis where $P^+$ and $P^\perp=0$ are diagonal.
As discussed in the Introduction, this is done
in SDLCQ by first discretizing the supercharge $Q^-$
and then constructing $P^-$ from the square of the supercharge:
$P^- = (Q^-)^2/\sqrt{2}$.
We have already introduced a finite discretization
in the transverse direction, characterized by the
compactification scale $L$ and cutoff or resolution $T$.
To complete the discretization of the supercharge, we introduce
discrete longitudinal momenta $k^+$ as fractions $nP^+/K$ of the
total longitudinal momentum $P^+$.  Here $n<K$ and $K$ are positive
integers.  The number of partons is also bounded by $K$.  The integer $K$
determines the resolution of the longitudinal discretization and is known
in DLCQ as the harmonic resolution~\cite{pb85}.
The remaining integrals in $Q^-$ are approximated by
a trapezoidal form.  The continuum limit in the longitudinal direction
is then recovered by taking the limit $K \rightarrow \infty$. 
In constructing the discrete approximation we drop the
longitudinal zero-momentum mode.
  
Our earliest SDLCQ calculations~\cite{Antonuccio:1999zu} were done using 
a code written for {\sc Mathematica} and performed on a PC. This code was 
rewritten in C++ for the work presented in~\cite{Haney:2000tk} and has been
substantially revised to reach higher resolutions~\cite{hpt2001}. The 
{\sc Mathematica} code continues to be useful for checking any changes 
in the C++ code. 

To obtain the spectrum of the CS theory we solve the 
complex eigenvalue problem, Eq.~(\ref{EVP}).
For the numerical evaluation we can exploit the structure of the supercharge 
\begin{equation}
Q^-=\left(\begin{array}{cc} 0 &A+iB\\ A^T-i B^T&0
\end{array}\right),
\end{equation}
where $A$ and $B$ are real matrices. The Hamiltonian has thus a simple 
decomposition into real and imaginary parts in the bosonic sector
\begin{equation}
P^-_{\rm boson}=AA^T+BB^T+i\left(BA^T-AB^T\right)\,,
\end{equation}
and the fermionic sector
\begin{equation}
P^-_{\rm fermion}=A^T A+B^T B+i\left(A^T B-B^T A\right)\,.
\end{equation}
We extract several of the lowest eigenstates by applying the Lanczos
algorithm~\cite{Lanczos}, as discussed in~\cite{hpt2001}.


\section{Setting the stage}  \label{sec:stage}

To analyze the full SYM-CS theory, it is convenient to first study three
subsets of the (2+1)-dimensional theory: 
the (1+1)-dimensional pure SYM theory ($T=0$, $\kappa=0$),
the pure SYM theory in three dimensions ($T>0$, $\kappa=0$), and 
the (1+1)-dimensional SYM-CS theory ($T=0$, $\kappa\neq 0$).
We have previously studied each of these theories
separately~\cite{Antonuccio:1998jg,Antonuccio:1998kz,hpt2001,CSSYM1+1}.

Let us start with the dimensionally reduced pure 
SYM theory~\cite{Antonuccio:1998jg,Antonuccio:1998kz}, for which
\begin{equation}
Q^-=gQ^-_{\rm SYM}(T=0)\,.
\end{equation}
There are two unique properties that characterize this theory. As we
increase the resolution, we find that there are new lower-mass states that
appear, and this sequence of states appears to accumulate at $M^2=0$.
In addition, there are massless BPS states. The dominant component of the
wave function of the BPS states can be arranged to have $2,3,\ldots,K$
particles, i.e.~up to the maximum number of particles allowed at
resolution $K$. Therefore at resolution $K$ there 
are $K-1$ bosonic BPS states and $K-1$ fermionic BPS states. 

Next we focus on the case $T>0$, $\kappa=0$, which is
the pure  SYM theory in $2+1$ dimensions~\cite{hpt2001}. 
The supercharge $Q^-$ becomes
\begin{equation}
Q^-=gQ^-_{\rm SYM}(T) + Q^-_\perp(T)\,.
\end{equation}
Again there are a number of unique properties that characterize the theory. At
small $g$ and low energy we recover the (1+1)-dimensional theory. At higher
energy we find a series of Kaluza--Klein states, which we discuss in detail
elsewhere~\cite{hpt02}.  At large coupling one might expect that the 
transverse $Q^-_\perp/g$ term would freeze out, and one would
see states that are a reflection of the (1+1)-dimensional theory. While we see 
these states, they are only a small part of the spectrum. $Q^-_{\rm SYM}$  
by itself wants to induce a large number of particles, as we discussed above. 
The contributions from $gQ^-_{\rm SYM}$ and $Q^-_\perp$ are therefore minimized
by a larger number of particles, each with a small transverse momentum.  We
therefore find that the average number of particles in the bound states grows 
with the coupling.  The massless BPS states of the (1+1)-dimensional SYM 
theory persist in 2+1 dimensions, but now the number of particles
in all of these states increases rapidly with the coupling. In Fig.~3 of 
Ref.~\cite{Antonuccio:1999zu} we displayed the average number of particles 
in the massless states as a function of the scaled coupling. At zero coupling 
we have the (1+1)-dimensional theory, and the states arrange themselves to have 
2 through $K$ particles. 
 
\begin{figure}
\begin{tabular}{cc}
\psfig{figure=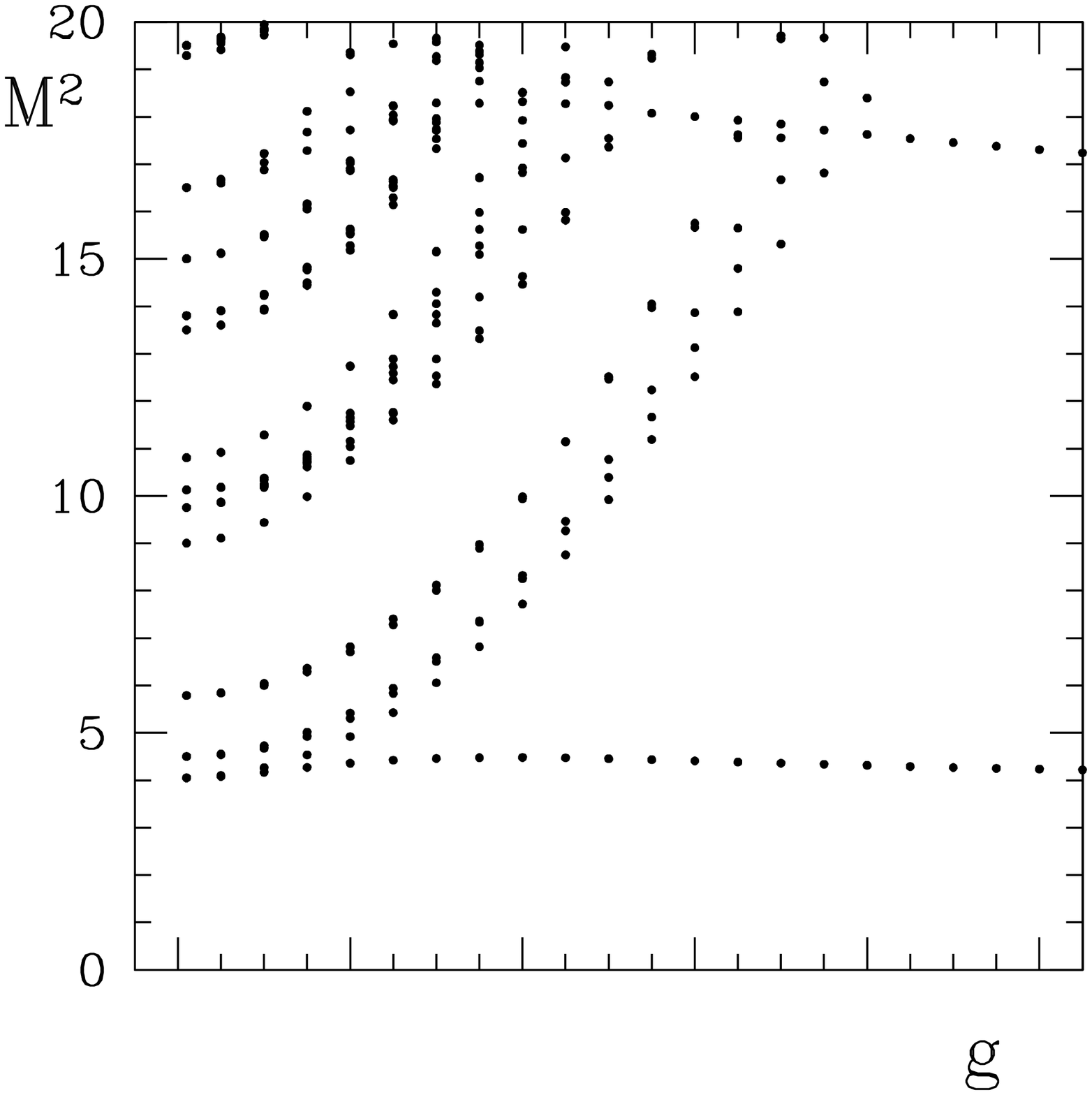,width=7.5cm,angle=0} &
\psfig{figure=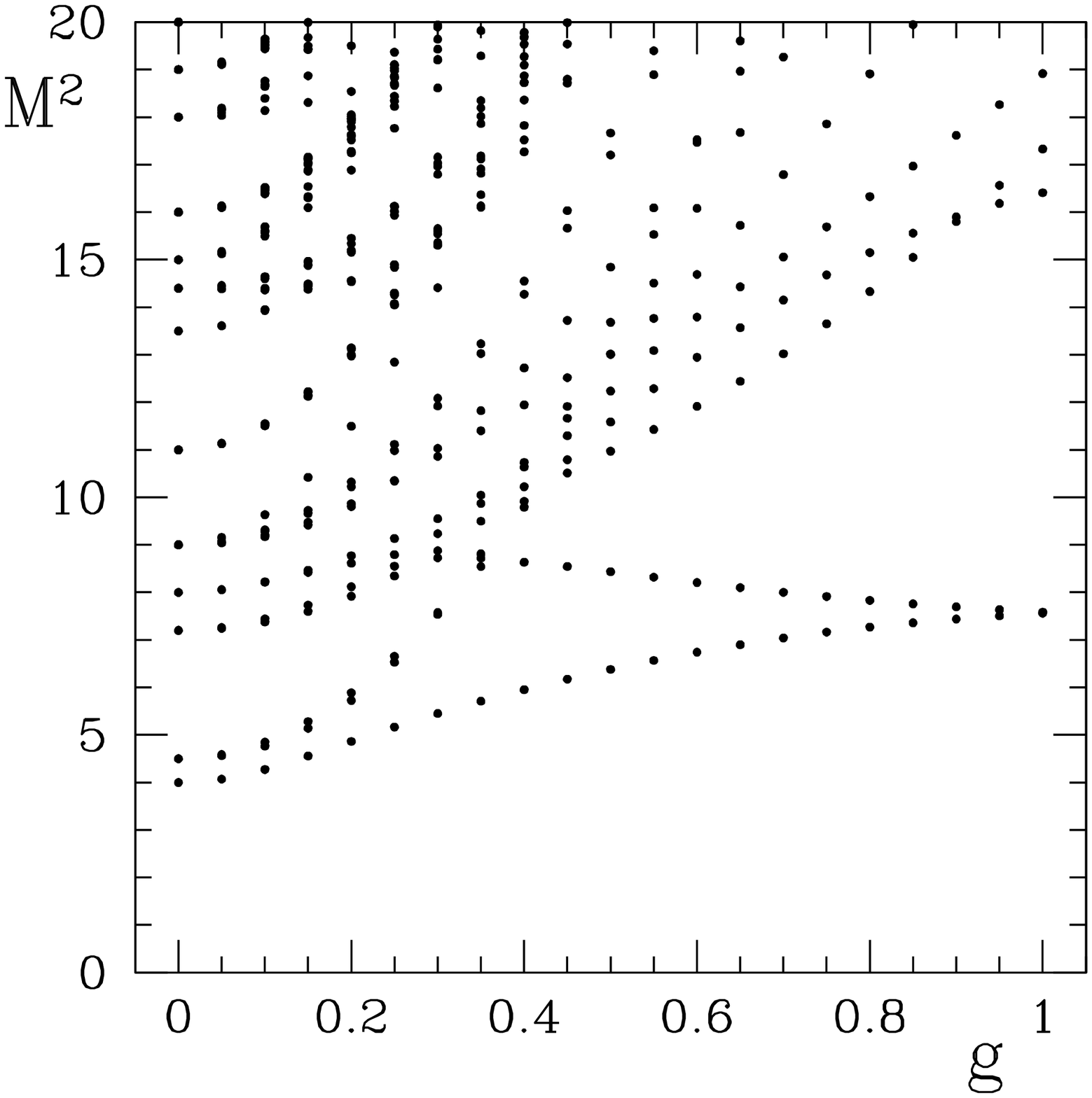,width=7.5cm,angle=0}\\
(a) & (b)
\end{tabular}
\caption{\label{OLD2}
The $Z_2$-even spectrum of SYM-CS theory as a function of the 
Yang--Mills coupling $g$ in (a) 1+1 dimensions and (b) 2+1 dimensions.
For (a), the longitudinal resolution is $K=9$, 
$M^2$ is measured in units of $\kappa^2$, 
and $g$ is in units of $\kappa\sqrt{\pi/N_c}$.
For (b), the longitudinal resolution is $K=6$, 
the transverse resolution is $T=1$, 
the Chern--Simons coupling is $\kappa=2\pi/L$, 
$M^2$ is in units of $4\pi^2/L^2$,
and $g$ is in units of $\sqrt{4\pi^3/N_c L}$.}
\end{figure}

If we set $T=0$ we get  SYM-CS theory reduced to two 
dimensions~\cite{CSSYM1+1}. Here we find that the most important
role of the CS term is to provide a mass for the constituents. This 
freezes out the long, lower-mass states that characterize (1+1)-dimensional 
SYM theory.  Interestingly, however, the massless BPS states become massive
approximate BPS states and have masses that are nearly independent of the 
YM coupling~\cite{BPS1+1}, as seen in Fig.~\ref{OLD2}(a) where we show
the two lowest $Z_2$-even approximate BPS states have squared masses
of $4\kappa^2$ and $16\kappa^2$ at infinite YM coupling. 

\vspace*{-0.5cm}

\section{Numerical Results}
\label{sec:results}

\subsection{DLCQ analysis}

\begin{figure}
\begin{tabular}{cc}
\psfig{figure=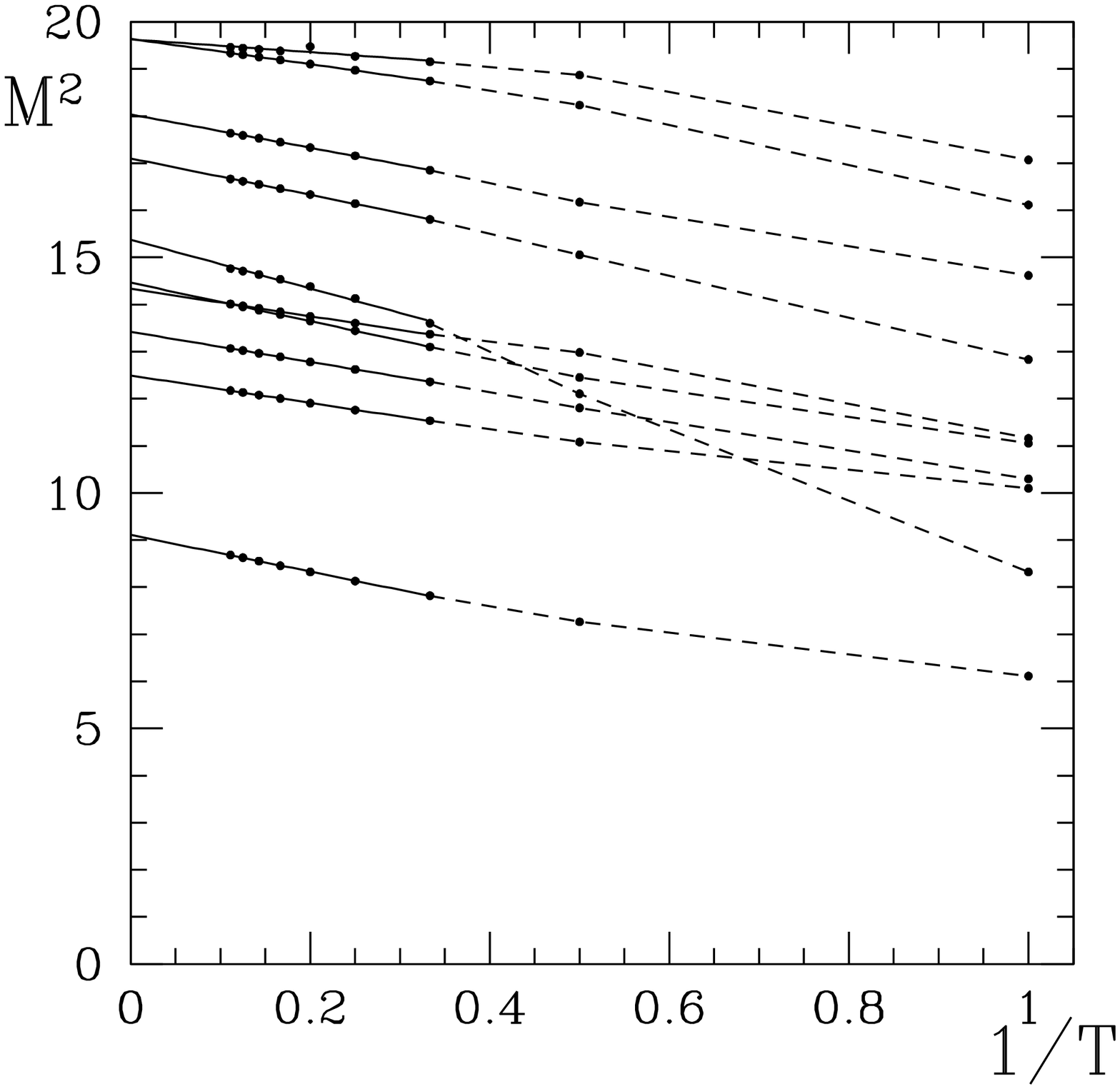,width=7.5cm,angle=0} &
\psfig{figure=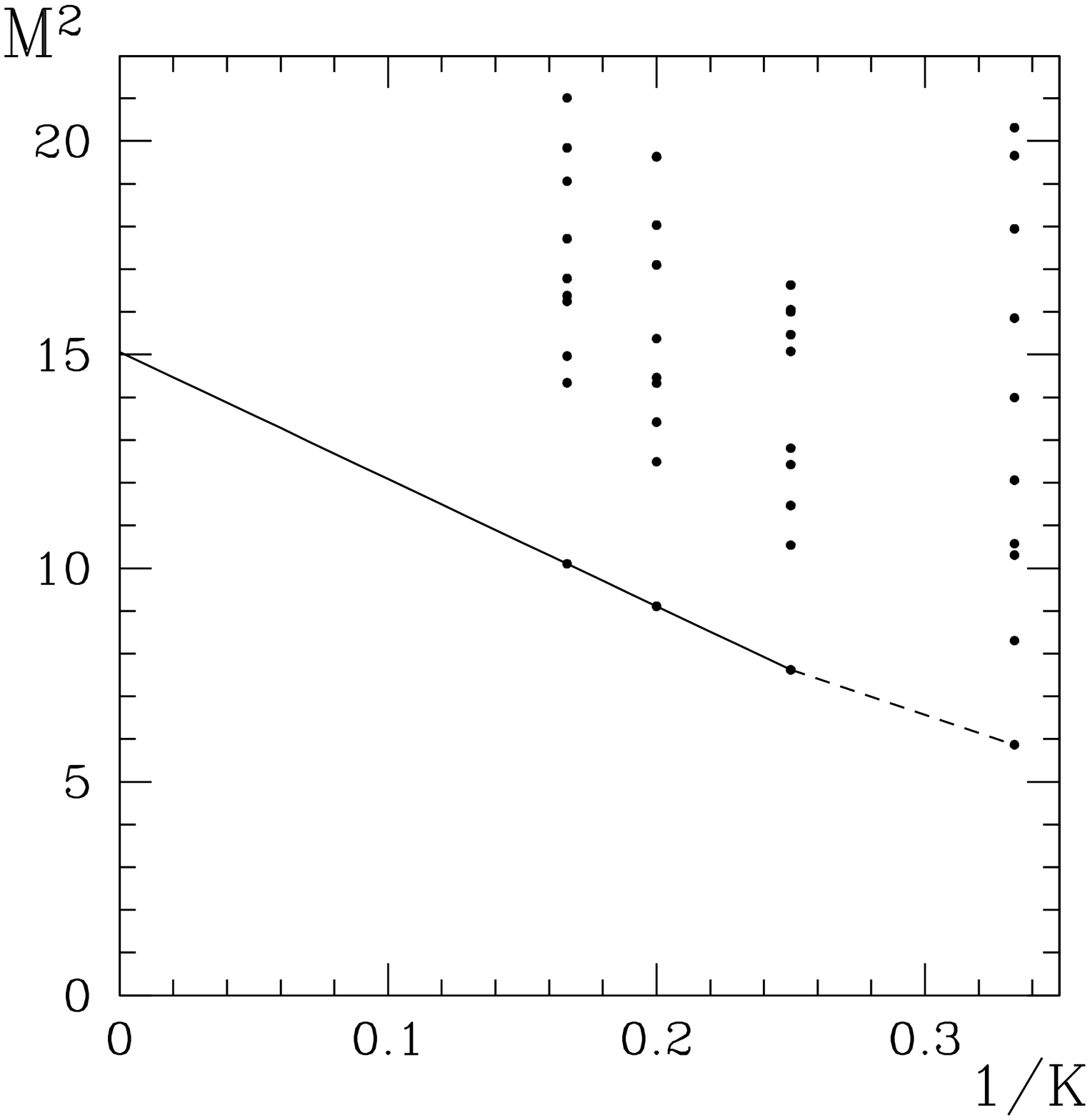,width=7.5cm,angle=0}\\
(a)&(b)
\end{tabular}
\caption{\label{g5} The $Z_2$-even spectrum of 
(2+1)-dimensional SYM-CS theory as a function of 
(a) transverse resolution $T$ and (b) longitudinal resolution $K$.
The values in (b) have been extrapolated to $T=\infty$.  The
coupling strengths are $g=0.5$, in units of $\sqrt{4\pi^3/N_c L}$, 
and $\kappa=2\pi/L$.} 
\end{figure}

We start the discussion of our new results with a standard DLCQ analysis 
of the bound states at\footnote{Values of $g$ are given in units
of $\sqrt{4\pi^3/N_c L}$.} $g=0.5$.  At each value of the resolution $K$ and 
in each $Z_2$ sector we look at the ten lowest-energy bound states at 
each transverse resolution $T$.  In order to extrapolate to infinite 
transverse momentum cutoff, we fit the curves of mass
squared as a function of the inverse transverse cutoff with a linear 
function in $1/T$, for $T \geq 3$, as shown in Fig.~\ref{g5}(a).
The intercepts of these curves are the transverse continuum results, 
which we plot as a function of $1/K$ in Fig.~\ref{g5}(b). Again we make 
a fit to the lowest mass state for $K \geq 4$, this time linear in $1/K$, 
and the intercept of this curve is the true continuum mass. We have not
attempted to fit the $1/K$ behavior of the higher states, because we are 
mainly interested in the lowest state, which is the analog of the lowest approximate BPS state in two dimensions.

For the case $g=0.1$ the average number of particles is well behaved,
in the sense that it stays small.  This allows for an additional 
approximation.  Namely, we can restrict the number of partons per Fock 
state.  This is a good approximation to the lowest ten states. Their 
average parton number is either two or three for even or odd $Z_2$, 
respectively.  This is clear since the states at small coupling are 
very close to the free states, which in turn are the discrete 
manifestations of the two and three particle continua starting at 
$M^2=4\kappa^2$ and $M^2=9 \kappa^2$, where $\kappa^2$ is 
the effective parton mass.  We restrict the number of partons in a state to
four.  This additional approximation allows us to carry the 
calculation to nine units of transverse resolution, $T=9$, and nine 
units of longitudinal resolution, $K=9$. The details of this analysis will
be presented elsewhere~\cite{hpt02}.  At $g=0.5$ we are unable to make this
approximation and are only able to go to maximum resolutions of $K=6$ and $T=4$.
The reason that this approximation breaks down is that, as we increase $T$ and 
$K$, the average number of partons in a state increases substantially. When 
this number approaches $K$, we miss a significant part of the 
wave function in the calculation.

In the $Z_2=-1$ bosonic sector the lowest state is a three-particle
state at weak coupling. All the analysis of this sector goes through 
similarly to the $Z_2=+1$ sector and will also be presented
elsewhere~\cite{hpt02}.

\subsection{Dependence of the spectrum on the Yang-Mills coupling}

In Fig.~\ref{OLD2}(b) we show the coupling dependence of SYM-CS
theory in 2+1 dimensions in the $Z_2=+1$ sector. We will restrict 
ourselves to this sector in the discussion that follows. For numerical 
convenience the results presented here are at specific and
low values of the longitudinal and transverse cutoffs $K$ and $T$ 
but at many values of the coupling. The point we want to make is that 
the phenomenon of approximate BPS states persists from two dimensions: 
some states appear to be nearly independent of the coupling at fixed 
$T$ and $K$. 

We can distinguish two regions in the low-lying spectrum. 
In the region of very small $g$ we have a spectrum of almost free 
particles, and in the regime of larger $g$ an approximate BPS state 
is clearly visible while the masses squared of the rest of the states 
grow like $g^2$. In the intermediate region we see a large
number of level crossings as the states rearrange themselves. This 
transition occurs around $g=0.3$. We will refer to couplings above 
this cross-over point as being strong, because the spectrum has 
completely rearranged itself here.

We have investigated the spectrum at
$g=0.1$ and $g=0.5$ as a function of $K$ and $T$
as typical representatives of weak and strong coupling, respectively.
It is obvious that some of the states seen in 
Fig.~\ref{OLD2}(b) move differently as a function of $K$ and $T$. 
At weak coupling we have the manifestation of a multi-particle continuum 
within a discrete approach, and the masses of the lowest states are very 
close together. As the coupling grows, a large gap develops 
between the lowest-mass state, which is an approximate BPS state, and the
other states in the spectrum, as seen in Fig.~\ref{OLD2}(b).  
The anomalous behavior of the next to lowest state, which  
is the only state to decrease in mass, is a small tranverse cutoff effect.
When $T$ is increased the gap is restored, as can be seen in Fig.~\ref{g5}(a):
the mass of the next to lowest state at $T=1$ increases rapidly as $T$ grows, 
leaving a mass gap in the transverse continuum limit. 
The computing time required to produce a graph of the spectrum as a function 
of the coupling at large $T$ is prohibitive, but we display the lowest masses 
in Table~\ref{gap} at $g=0.1$ and $g=0.5$. We see that at
$g=0.1$ the lowest 10 states are more or less uniformly spaced from 
$M^2=4$ to $M^2=10$ in units of $4\pi^2/L^2$. This is what we would 
expect since at small coupling these states are closely related to the 
two-particle continuum states. At strong coupling, $g=0.5$, we see that a 
large gap has developed between the state at $M^2=8.93$ and $M^2=13.14$. 
These results are obtained at resolution $K=6$, but from Fig.~\ref{g5}(b) 
it appears that this gap is growing with the longitudinal resolution, and, 
therefore, in the continuum we expect it will be even larger. 

We have studied the average number of particles of all the states in all 
of the sectors and a detailed discussion will be presented 
elsewhere~\cite{hpt02}. Here it is interesting to note that for the 
state at $M^2=8.93$ we find an average parton number
squared $\langle n^2 \rangle =8.23$. This is close  
to the expected value based on the duality discussed in SYM-CS in 1+1
dimensions~\cite{BPS1+1}, where we found that the mass at infinite coupling 
is the average parton number squared at CS coupling unity. 
By contrast, the average parton number squared
$\langle n^2 \rangle =9.9$ of the second lightest state is far away from its 
mass squared $M^2=13.14$.

\begin{table}
\centerline{
\begin{tabular}{|c||c|c|c|c|c|c|c|c|c|c|}\hline
$g$ & \multicolumn{10}{|c|}{$M^2$}\\\hline
$0.1$ & 4.36 & 4.85 & 4.93 & 7.41 & 7.47 & 8.35 & 8.36 & 9.34 & 9.43 &  9.48\\ 
\hline
$0.5$ & 8.93 & 13.14 & 13.88 & 14.92 & 15.10 & 15.62 & 15.82  & 17.38 & 18.35
&19.76\\ 
\hline
\end{tabular}}
\caption{\label{gap}
Mass squared in units of $4\pi^2/L^2$ of the lowest-mass states 
at $g=0.1$ and $g=0.5$, in units of $\sqrt{4\pi^3/N_c L}$, 
showing the development of a gap above the lowest 
state. Results in both cases are for resolutions $K=6$ and $T=4$ and for
Chern--Simons coupling $\kappa=2\pi/L$.}
\end{table}

\subsection{Structure Functions}

In this section we will present a discussion of the structure functions 
of the approximate BPS states.  A complete discussion of the structure 
functions of all the states will be presented elsewhere~\cite{hpt02}. 
The structure functions are functions of both the longitudinal and 
transverse momentum and are defined to be the probability that a particle 
of type $A$ will have a longitudinal momentum fraction $x=k^+/P^+$ and 
transverse momentum $k^\perp$. It is given as follows in terms of
the light-cone wave function $\psi$:
\begin{eqnarray}
\hat{g}_A(x,k^\perp)&=&\sum_q\int_0^1 dx_1\cdots dx_q \int_{-\infty}^{\infty}
dk^\perp_1\cdots dk^\perp_q 
\delta\left(\sum_{i=1}^q x_i-1\right)
\delta\left(\sum_{j=1}^q k^{\perp}_j\right)\nonumber \\
&&\qquad\times
\sum_{l=1}^q \delta(x_l-x)\delta(k^\perp_l-k^\perp)\delta^A_{A_l}
|\psi(x_1,k^\perp_1;\ldots x_q,k^\perp_q)|^2\,.
\end{eqnarray}
We present the discrete version of the structure function as a function of
the number of units of longitudinal momentum and the number of units of 
transverse momentum.  No attempt has been made to extrapolate these curves.

\begin{figure}
\begin{tabular}{cc}
\psfig{figure=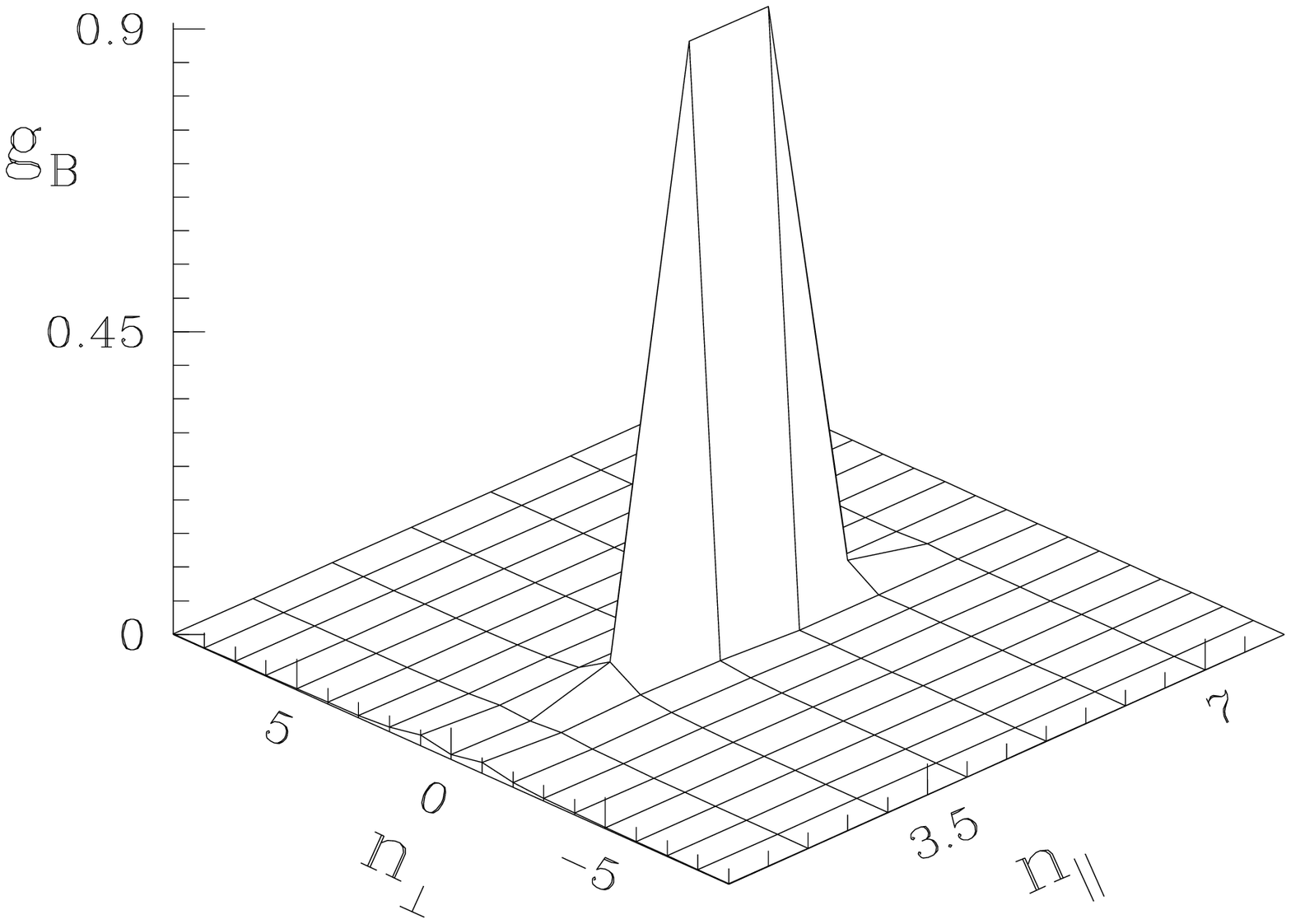,width=8cm,angle=0}&
\psfig{figure=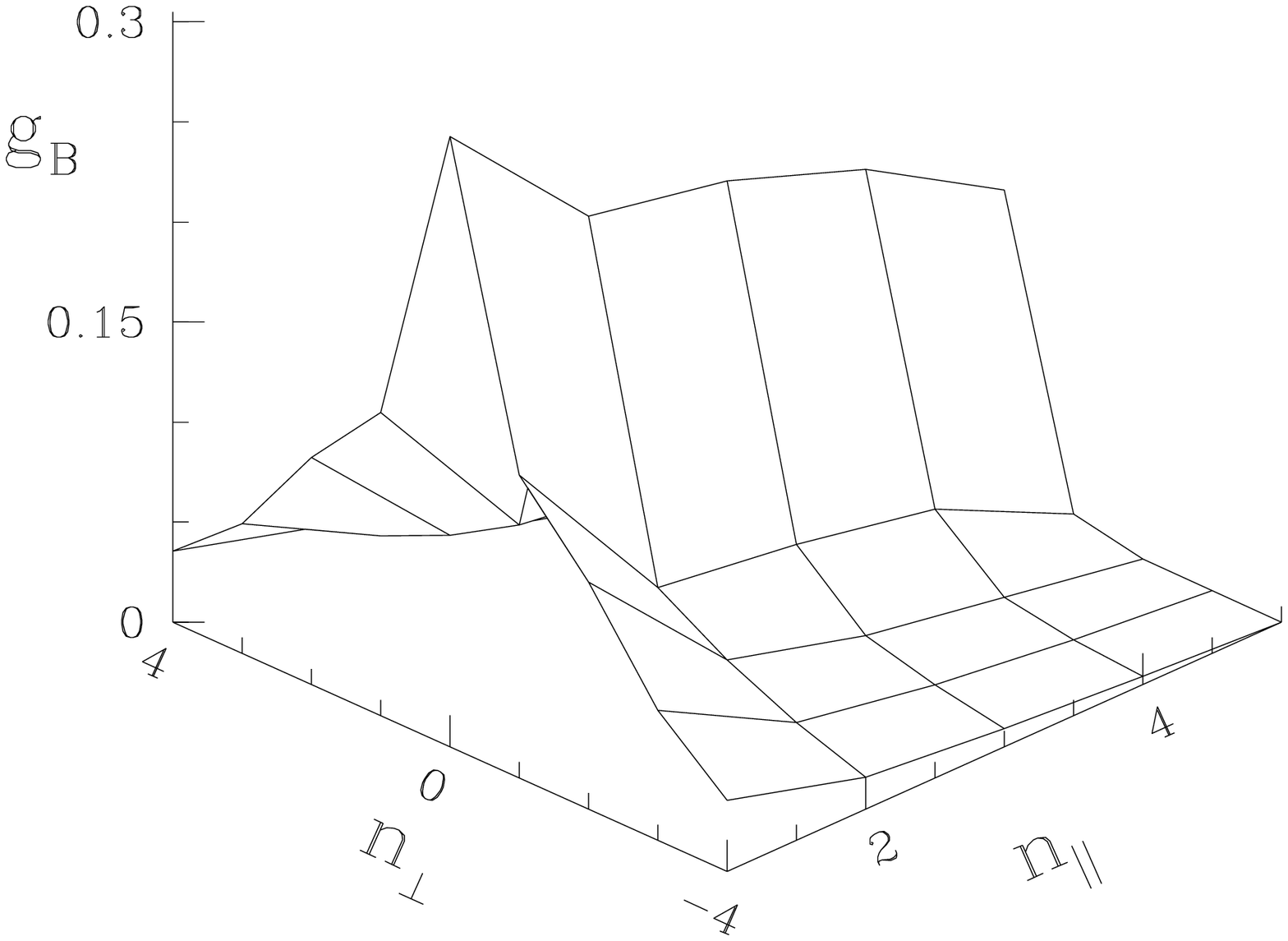,width=8cm,angle=0}\\
(a) & (b)
\end{tabular}
\caption{\label{structure}
Structure function of the lowest approximate BPS state at 
(a) $g=0.1$ and (b) $g=0.5$, in units of $\sqrt{4\pi^3/N_c L}$.  
The numerical resolutions are $K=6$ and $T=4$.  
The value of the Chern--Simons coupling is $\kappa=2\pi/L$.} 
\end{figure}

In Fig.~\ref{structure} we see the structure function of the approximate BPS
state of the SYM-CS theory for $g=0.1$ and $g=0.5$.  At weak coupling it has
one localized peak in both variables. In fact this is true of all of the 
low-mass states of this theory at weak coupling, and the approximate BPS
state does not appear in any way special. As we move to stronger coupling,
however, all structure functions of the normal bound states remain peaked in
$n_{||}$, while the peak moves down to lower values of $n_{||}$ as the 
number of particles increases.  In contrast we see that the BPS state is 
nearly flat in $n_{||}$, although it has a small peak at $n_{||}=1$. 
This behavior is unique to this state. 

At this point we do not have an understanding of why the approximate BPS 
state has this behavior. It is, however, clearly important that it does, if 
for no other reason than to give one a means to identify such a state 
experimentally. 

\section{Outlook} 
\label{sec:Outlook}

In our earlier work on dimensionally reduced SYM-CS 
theory~\cite{CSSYM1+1,BPS1+1}, we found that there were
states whose properties are largely dictated by the properties 
of the massless BPS states of the underlying SYM theory.  We saw 
that the masses of these states were nearly constant as functions
of the YM coupling, and we therefore called these states approximate 
BPS states. The mass scale of these states is set by the CS coupling 
and the average number of partons. The binding energy is approximately 
zero, because of the underlying properties of the BPS state.

In this paper we have looked for the effect of these approximate BPS 
states in the full (2+1)-dimensional SYM-CS theory. We have looked at 
the low mass spectrum and found that at strong coupling there are states 
with an anomalously low mass. The mass scale for these states is set by the 
CS coupling and the average number of partons, similar to the (1+1)-dimensional
case. These states are a reflection of the exact BPS states that exist 
in (2+1)-dimensional SYM theory. In addition, we examined the 
structure functions for these states.  We find that they are 
different from the structure functions of the other states; they are 
nearly flat in longitudinal momentum. 

It appears that we have found a new mechanism for generating low-mass
bound states. There are two issues, however, that need to be addressed 
before speculating on the possibility that this mechanism might be 
realized in nature.  First, we have not included any matter fields in 
this theory.  There has been some work done with matter fields in
SDLCQ~\cite{Lunin:2001im}, and one of our next projects will be to include 
them.  Second, we have not included supersymmetry breaking.
Instead, we added masses for the adjoint fields by including a CS 
term. The calculational technique we use, SDLCQ, relies heavily on the 
fact that we have an exactly supersymmetric theory, and it is essential,
therefore, that we introduce masses without breaking the supersymmetry. 
Inclusion of supersymmetry breaking remains an outstanding problem in 
this numerical approach.

\section{Acknowledgments}
This work was supported in part by grants of computing time
from the Minnesota Supercomputing Institute and by the U.S. 
Department of Energy. 

\end{document}